\pdfoutput=1
\documentclass[10pt]{article}
\usepackage{times}
\usepackage{fullpage}
\usepackage{algorithm, algorithmic}
\usepackage{amsmath, amssymb, amsthm}
\usepackage{graphicx}
\usepackage{algorithm, algorithmic}
\usepackage{color}

  {\begin{description}%
    \setlength{\itemsep}{1pt}%
    \setlength{\parskip}{1pt}}%
  {\end{description}}

  {\begin{enumerate}%
    \setlength{\itemsep}{1pt}%
    \setlength{\parskip}{1pt}}%
  {\end{enumerate}}

  {\begin{itemize}%
    \setlength{\itemsep}{1pt}%
    \setlength{\parskip}{1pt}}%
  {\end{itemize}}

\begin{document}

\title{Back to the Future: an Even More \\
Nearly Optimal Cardinality Estimation Algorithm}


\author{Kevin J. Lang \\ Oath Research}

\date{August 22, 2017}


\maketitle

\begin{abstract}
We describe a new cardinality estimation algorithm that is extremely space-efficient. 
It applies one of three novel estimators to the compressed state of the Flajolet-Martin-85 
coupon collection process. In an apples-to-apples empirical comparison against compressed 
HyperLogLog sketches, the new algorithm simultaneously wins on all three dimensions of the 
time/space/accuracy tradeoff. Our prototype uses the zstd compression library, and produces sketches 
that are smaller than the entropy of HLL, so no possible implementation of compressed HLL can match 
its space efficiency. The paper's technical contributions include analyses and simulations of the
three new estimators, accurate values for the entropies of FM85 and HLL,
and a non-trivial method for estimating a double asymptotic limit via simulation.
\end{abstract}

\section{Introduction}

In the cardinality estimation task, an algorithm must process a multiset of identifiers
that is much larger than the amount of memory that the algorithm is allowed to use.
The identifiers are processed in a streaming fashion, i.e. one at a time. 
At the end of the stream, the algorithm must estimate the number of {\em distinct} identifiers
in the multiset.
This task is ubiquitous in the internet and big-data industries. To give just one example,
it could be useful to know how many unique IPv6 addresses appear in a year's worth of logs
from a million servers.

HyperLogLog \cite{hllpaper} sketches have a well-deserved reputation for being the best solution when this
task must be accomplished in a way that is space efficient.
Although HLL is often used in its uncompressed form
requiring $O(\log \log n)$ bits per row,
\cite{frenchthesis} proved that on average, 
HLL sketches can be compressed to 
less than 3.01 bits per row  --- a constant that is 
independent of $n$, thus providing space optimality (to within a constant factor)
in an algorithm that is more practical than the theoretical sketches of \cite{knwpaper}.

Interestingly, HLL can be viewed as a lossily compressed version of its 
historical predecessor FM85 \cite{fm85paper}.
We report the surprising discovery that the information which
is discarded by this lossy compression can more than pay for itself
when it is retained.

\subsection{Preliminaries}

Throughout this paper, the symbol $n$ denotes the number of distinct identifiers in the stream.
The symbol $k$ denotes a parameter that controls the accuracy and space
usage of the sketches. The term Standard Error means (Root Mean Squared
Error) / $n$. The term Error Constant means
$\lim_{k\rightarrow\infty} (\lim_{n\rightarrow\infty} (\sqrt{k} \times \mathrm{RMSE} / n)).$

All of the sketches that we will discuss are based on stochastic
processes that are driven by random draws from probability distributions. 
The processes can be simulated using random number generators, 
but in an actual system the random numbers
are replaced by high quality hashes of the identifiers.
The hashes provide repeatable randomness, and if the sketch update rule is idempotent,
they transform a scheme for counting into a scheme for {\em distinct} counting.
This transformation explains why the rest of this paper is focused on 
stochastic processes, and why it discusses counting without mentioning distinctness.

\begin{figure}
\begin{center}
\begin{tabular}{|l|l|l|l|l|}
\hline
 & \multicolumn{2}{|c|}{Existing Error} & Space & Novel Error \\
 & \multicolumn{2}{|c|}{Constant for}   & Efficiency   & Constant for \\
Estimator Category & \multicolumn{2}{|c|}{HLL Sketches}   & Threshold    & FM85 Sketches \\
\hline
Best Known Summary Statistic   & \cite{hllpaper} & 1.0389618 & 0.807 & 0.6931472 \\ 
Minimum Description Length     &                 & 1.037     & 0.805 & 0.649 \\
Historic Inverse Probability   & \cite{tingHIP}  & 0.8325546 & 0.646 & 0.5887050 \\ 
\hline
\end{tabular}
\caption{The low Error Constants in the right column
show that all three of our novel estimators cause FM85 sketches
to have better accuracy per bit of entropy than HLL sketches.}
\label{top-level-comparison-figure}
\end{center}
\end{figure}

\subsection{Overview of Results}

We begin by calculating the asymptotic (in $n$) entropy of FM85 and HLL sketches,
which turns out to be $(4.70 \times k)$ bits for FM85,
and $(2.83 \times k)$ bits for HLL.
Because the asymptotic Standard Error of each type of sketch is a constant
divided by $\sqrt{k}$, these entropy values imply that
compressed FM85 can be more space-efficient than compressed 
HLL provided that (FM85 Error Constant) / (HLL Error Constant) 
$\;< \sqrt{2.83/4.70} \approx 0.776$.

Figure~\ref{top-level-comparison-figure} tabulates: 1) the already-known
Error Constants for three HLL estimators; 2) threshold values 
that are 0.776 times the HLL constants; 3) the Error Constants
for this paper's three novel FM85 estimators. All three of our new estimators cause FM85 sketches
to have better accuracy per bit of entropy than HLL sketches.

We note that the constant for the original FM85 estimator was $\sim 0.78$. 
This was already below the threshold of 0.807, but the 0.693 of the current paper's ICON estimator renders 
the original estimator obsolete.


\subsection{Contributions}

\begin{itemize}
\item The surprising discovery that FM85 sketches are more accurate per bit
of entropy than HLL sketches.
\item Three novel estimators for the FM85 sketch, together with analyses and simulations
showing that all of them are more accurate than the estimator from the original paper.
\item An apples-to-apples comparison of implementations of compressed HLL and compressed FM85,
in which compressed FM85 simultaneously wins on all 3 dimensions of the time/space/accuracy tradeoff.
\item The most precise calculations of the entropy of FM85 and HLL sketches to date.
\item A non-trivial method for estimating the value of the 
double asymptotic limit which defines Error Constants.
\end{itemize}

\section{Approximate Counting via Coupon Collection}

Although our results pertain to FM85 sketches specifically,
our definitions and estimators apply to any counting sketch that is
based on the stochastic process of coupon collecting. The probabilities of the coupons need not be equal,
but they do need to sum to 1. 
Let $p_i$ denote the probability that coupon $i$ is collected on a
given draw.
Let $b^n_i$ denote the Bernoulli random variable indicating that
coupon $i$ has been collected at least once during the current run of $n$ draws.  
Let $q^n_i$ denote the probability 
that coupon $i$ has been collected at least once during the current run of $n$ draws,
and note that $q^n_i = 1 - (1 - p_i)^n$.

\subsection{Entropy:}


Consider an infinite number of repetitions of a sequence of $n$ draws
from a given set of coupons. The resulting sets of collected coupons
are then compressed by arithmetic coding \cite{wittencoding} using the probabilities
$q^n_i$. This causes the average compressed size of the sets to equal
their entropy, which is
\begin{equation}
\mathrm{Entropy} = \sum_i -q^n_i \log_2 q^n_i - (1 - q^n_i) \log_2 (1 - q^n_i)
\end{equation}
\noindent When a specific set of collected coupons has been compressed using this method, 
its size in bits is:
\begin{equation} \label{ant1}
\mathrm{Space} = \sum_i -b^n_i \log_2 q^n_i - (1 - b^n_i) \log_2 (1 - q^n_i)
\end{equation}

\subsection{MDL Estimator:}

In practice we don't know the value of $n$, but
formula (\ref{ant1}) can be put to good use as the foundation
for a Minimum Description Length estimator \cite{mdlpaper}
that maps a concrete set of collected coupons to an estimate of $n$:

\begin{equation}\label{ant2}
\hat{N}_{\mathrm{MDL}} = \arg \min_m 
\sum_i -b^n_i \log_2 q^m_i - (1 - b^n_i) \log_2 (1 - q^m_i)
\end{equation}

\noindent $\hat{N}_{\mathrm{MDL}}$ can be computed via binary search over guesses
$m$ of the value of $n$. 
The downside of this estimator is that every step of the binary search
requires formula (\ref{ant2}) to be
evaluated over the entire set of collectible coupons.

\subsection{ICON Estimator:}

While not a sufficient statistic, the number of collected coupons 
$C$ is a better summary statistic than many authors have realized.
It can be mapped to an estimate in several ways, including the
following which outputs the $n$ that causes the expected number 
of collected coupons to match the number that have actually been collected.

\begin{equation}
\hat{N}_{\mathrm{ICON}}
= \arg \min_m (C - \sum_i q^m_i)^2.
\end{equation}

\noindent As with the MDL estimator, the ICON estimator
could be evaluated at query time via binary search,
but since it's a function of the single integer-valued quantity $C$,
the mapping can be pre-computed and stored in a lookup table,
after which the cost of producing an estimate is a single cache miss.
[The name ``ICON'' is a loose acronym for ``Inverted N to C mapping'',
where N is the number of unique identifiers in the stream, and C is
the number of collected coupons.]

\subsection{HIP Estimator:}

The Historic Inverse Probability estimator \cite{cohenHIP,tingHIP}
can be implemented with two variables that are 
maintained incrementally:
an accumulator $A$ that starts at 0.0, and a remaining
probability $R$ that starts at 1.0.
Whenever a novel coupon $i$ is collected, the accumulator is 
updated by the rule $A \leftarrow A + (1/R)$, then the probability
is updated by the rule $R \leftarrow R - p_i$. 
The estimator $\hat{N}_{\mathrm{HIP}}$ is the current value of $A$.

\subsection{Mergeability:}

These sketches can be merged by unioning their sets of collected coupons. 
A sketch produced by merging two sketches is identical to the sketch
that would result if the original streams had been concatenated then
processed by a single sketch. As a result, estimates from the
merged sketch and the single sketch are the same when 
the ICON or MDL estimator is used.
However, HIP estimators depend on the order in which the stream was processed, 
and cannot be re-calculated from the information in the sketch. As a 
result they do not survive merging, limiting their usefulness
in the massively parallel systems employed by industry.

\subsection{Instantiations:}

The above definitions apply to an infinite family of counting sketches, each
specified by a different
rule for assigning probabilities to coupons.
Numerous members of this family have been proposed and studied before;
prominent examples include the Linear Time Counting sketch of \cite{whangpaper}
which employs $k$ equiprobable coupons, and the cardinality estimation sketch 
of Flajolet-Martin-85 \cite{fm85paper}. The coupons of the FM85 sketch will be described 
in the next section.


\section{Entropy of FM85 Sketches}

Abstractly if not concretely, the FM85 data structure is a rectangular
matrix of cells, each associated with a collectible coupon and a boolean
state (representable by a 1 or 0) that indicates whether the cell's
coupon has been collected yet. The matrix has $k$ rows and an infinite number
of columns. From left to right, the columns of the matrix are
numbered $1 \le j < \infty$, and the single-draw probability of a coupon in column
$j$ is $1/(k \cdot 2^j)$. 

\begin{figure}
\begin{center}
\includegraphics[width=0.3\linewidth]{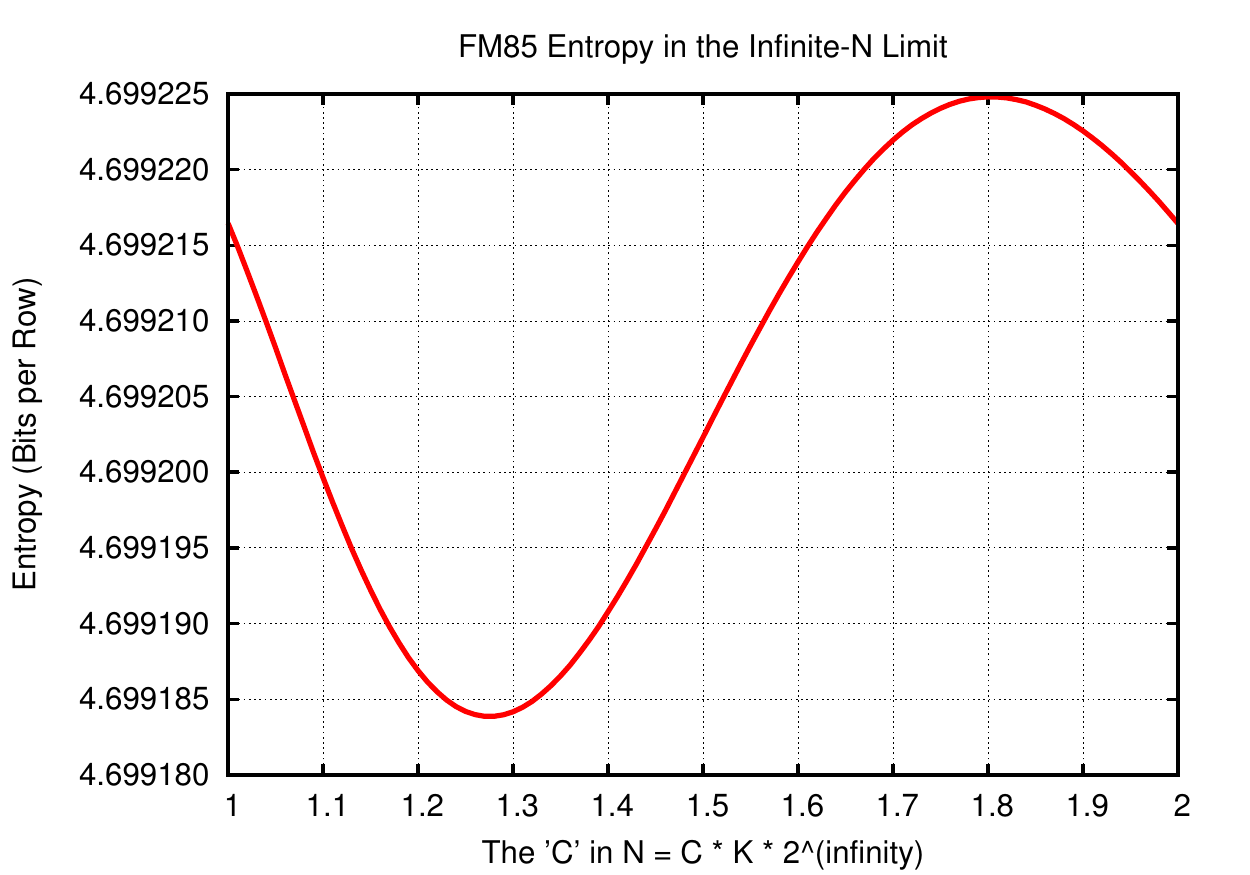}\quad 
\includegraphics[width=0.3\linewidth]{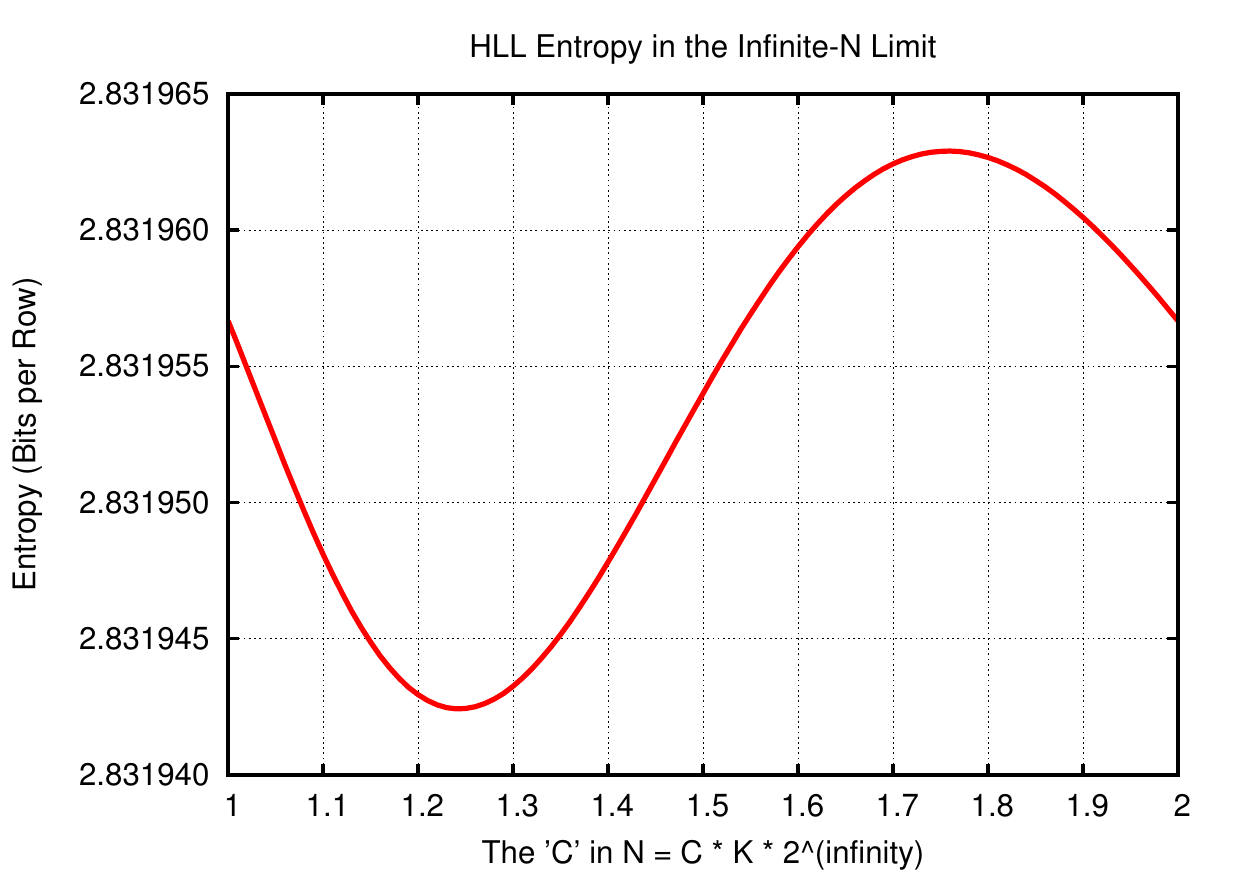}
\caption{Upper bounds on the asymptotic entropy of FM85 and HLL.}
\label{oscillation-figure}
\end{center}
\end{figure}

To calculate the asymptotic entropy of FM85, we begin by 
assuming that $n$ is a number
that can be factored as $n = c k 2^b$, where $1.0 \le c < 2.0$, and 
$b$ is a very large integer.
Consider a cell for which $j = b+d$.
The probability of the cell's coupon not being collected during 
a sequence of $n$ draws is
\begin{equation}\label{r-j-definition}
r_d = \left(1 - \frac{1}{k2^{b+d}}\right)^{c k 2^b} 
\substack{\longrightarrow \\ b \rightarrow \infty}
\left(\frac{1}{e}\right)^{\frac{ck2^b}{k2^{b+d}}}
=\;\;
\left(\frac{1}{e}\right)^{c2^{-d}}
\end{equation}
With $q_d = 1 - r_d$ denoting the probability that the coupon has been collected,
the entropy associated with the cell is
\begin{equation}\label{FM85-coupon-entropy}
\mathrm{FM85CellEntropy}_d \;=\; r_d \log_2 (1 / r_d) + q_d \log_2 (1/q_d)
\end{equation}
\noindent The overall entropy of the sketch is 
\begin{equation}\label{FM85-total-entropy}
\mathrm{FM85TotalEntropy} \;=\; k \!\! \sum_{d=-\infty}^{\infty} \mathrm{FM85CellEntropy}_d.
\end{equation}
Using the $(1/e)^{c2^{-d}}$ asymptotic value of $r_d$, we evaluated this formula 
numerically 
and found that the per-row entropy is the oscillating
function of $\log c$ which is depicted in Figure~\ref{oscillation-figure} (left).
Fourier analysis reveals that this function is the constant
4.699204337 plus a pure sine wave of tiny 
amplitude.\footnote{The analyses in \cite{fm85paper,hllpaper,frenchthesis}
all found oscillations in the asymptotic behavior of quantities associated with FM85 and HLL sketches.}

We remark that by summing the cell's entropies, we have implicitly
assumed that the random variables associated with their boolean
states are independent conditioned on $n$.
This isn't quite true, but independence increases the entropy of a composite system,
so the result of this calculation is an upper bound on the sketch's true 
entropy. 
The full version of this paper includes a Monte Carlo lower bound 
that differs from the upper bound by 1/10,000 of a bit.

\section{Entropy of HLL Sketches}

Having already worked out the entropy of each cell in the FM85 coupon matrix,
there is an easy way to determine the entropy of the HLL data structure.
The key insight is that an HLL sketch can be viewed as an FM85 sketch
that has discarded all information about any cell that lies to the left
of the rightmost collected coupon in each row.

This implies that the entropy of the HLL data structure is
the sum over all cells of the FM85 matrix of the following quantity: 
(the FM85 entropy of the cell) $\times$ 
(the probability that HLL has not yet discarded the cell's information).
The latter quantity is the
probability that no coupon to the right of the cell has been
collected, which by a simple argument is equal to the probability that
the cell's own coupon has not been collected, which is the quantity
$r_d$ defined by (\ref{r-j-definition}).
Therefore the HLL entropy formula is the FM85 formula (\ref{FM85-total-entropy})
with each cell's term multiplied by $r_d$, in other words:
\begin{equation}\label{HLL-total-entropy}
\mathrm{HLLTotalEntropy} \;=\; k \!\! \sum_{d=-\infty}^{\infty} r_d \cdot \mathrm{FM85CellEntropy}_d.
\end{equation}

\noindent Using the $(1/e)^{c2^{-d}}$ asymptotic value of $r_d$, we evaluated this formula 
numerically 
and found that the per-row entropy is the oscillating
function of $\log c$ depicted in Figure~\ref{oscillation-figure} (right).
Fourier analysis reveals that this function is the constant
2.831952664 plus a pure sine wave of tiny amplitude. 
This is technically an upper bound, but based on empirical evidence
that will be presented in the full paper, and also on the fact 
that this value is only an upper bound because of the 
asymptotically nonexistent dependence between cell states,
we believe that it is essentially tight.

\section{ICON Estimator for FM85 Sketches}

The FM85 stochastic process maps values of $n$, the true number of unique
items in the stream, to values of the random variable $C$ which represents
the current number of collected coupons.
Given $n$ and $k$, the expected value of $C$ is
\noindent 
\begin{equation}
E(C) \;=\; \sum_i q^n_i \;=\; k \sum_j \; q_j^n, \hspace{12em} \label{c-mean-formula}
\end{equation}
\noindent where $1 \le j < \infty$ ranges over the sketch's columns,
and $q^n_j = 1 - (1 - \frac{1}{k 2^j})^n$.

\vspace{0.5em}
\noindent For any fixed $k$, formula~(\ref{c-mean-formula}) is a 
mapping from $n$ to $E(C)$ whose functional inverse (viewed as a mapping
from $C$ to $\hat{n}$) defines the ICON estimator. By wrapping a 
binary search around formula~(\ref{c-mean-formula}), this
inverse can be calculated 
for every value
of $C$ and stored in a lookup table, after which the cost of producing an estimate is a single cache miss.

\subsection{Informal Error Analysis}

\noindent Given $n$ and $k$, the variance of $C$ is
\begin{eqnarray}
\sigma^2(C) 
\;\; = & \sum_i q^n_i \; (1 - q^n_i) 
\;\; = \;\; k \; \sum_j \; q^n_j \; (1 - q^n_j). \label{c-var-formula}
\end{eqnarray}
In Appendix A, we prove that when $k$ is large and $n >> k$, 
\begin{eqnarray}
\sigma^2(C) \;\; \approx &k. \hspace{16em} \label{c-var-approx}
\end{eqnarray}
\noindent The exact formula (\ref{c-mean-formula}) for the expected value of $C$ is
hard to analyze, so instead we analyze the following approximation 
to its asymptotic behavior that can be 
obtained from  an informal symmetry argument.
\begin{equation}
E(C) \;\; \approx \;\; f(n) = \frac{k}{L} \; \ln \frac{n}{Dk}, 
\quad \mathrm{with} \; L = \ln 2, \;\; \mathrm{and} \; D \approx 0.7940236
\label{c-mean-approx}
\end{equation}
\noindent The ICON estimator can then be approximated as follows.\footnote{
When $n >> k$, the ratio (exact ICON estimate) / (approximate ICON estimate) 
is the constant 1.0 plus a tiny oscillation that goes through one cycle each time
that $n$ doubles.}
\begin{equation}\label{icon-approximation}
\hat{N}_{\mathrm{ICON}} 
\;\; \approx \;\; f^{-1}(C) 
= D k \; \exp \left(\frac{LC}{k} \right).
\end{equation}

\noindent Assuming that the distribution of $C$ is well-approximated by a
Gaussian with mean $\mu$ and variance $\sigma^2$, (\ref{icon-approximation}) implies
that the ICON estimator has a log-normal distribution with 
mean $\exp(\mu + \sigma^2/2)$ and variance
$[\exp(\sigma^2) - 1] \exp (2\mu + \sigma^2)$.
Therefore, via some algebra that is omitted to save space,
(\ref{c-var-approx}) and
(\ref{c-mean-approx}) imply
\begin{eqnarray}
\mathrm{Bias}(\hat{N}_{\mathrm{ICON}}) \;\approx & n \; [ \exp (L^2/(2k)) - 1) ] \\
\sigma^2(\hat{N}_{\mathrm{ICON}})      \;\approx & n^2 \; [\exp(2L^2/k) - \exp(L^2/k)] \\
\mathrm{MSE}(\hat{N}_{\mathrm{ICON}}) \; \approx  & n^2 \;[\; \exp(2L^2/k) - 2 \exp(L^2/2k) + 1 \;].
\end{eqnarray}
\noindent By plugging in the Maclaurin series for exp(),
we can recover the leading terms of these formulas
\begin{eqnarray}
\mathrm{Bias}(\hat{N}_{\mathrm{ICON}})/n    
\;\;\approx & \frac{(\ln 2)^2}{2k}
\;\;\approx & \frac{0.24022650}{k} \\
\mathrm{\sigma}(\hat{N}_{\mathrm{ICON}})/n  
\;\;\approx & \frac{\ln 2}{\sqrt{k}}
\;\;\approx &  \frac{0.69314718}{\sqrt{k}} \\
\mathrm{RMSE}(\hat{N}_{\mathrm{ICON}})/n    
\;\;\approx & \frac{\ln 2}{\sqrt{k}}
\;\;\approx & \frac{0.69314718}{\sqrt{k}}. \label{fm85-icon-rmse}
\end{eqnarray}

\vspace{0.5em} 
\noindent Because the constant in (\ref{fm85-icon-rmse}) matches our simulations to 4.5 decimal digits
(the noise floor of the empirical measurements), 
we conjecture that a more rigorous analysis of the FM85 ICON estimator would arrive at essentially the same result.

\section{HIP Estimator for FM85 Sketches}


Based on an informal argument that is similar in spirit to the theory
of self-similar area cutting processes in \cite{tingHIP}, we conjecture that in the 
asymptotic limit, the variance of the FM85 HIP estimator is
\begin{equation} \label{conjectured-hip-variance}
\sigma^2(\hat{N}_{\mathrm{HIP}}) \;\; \approx \;\; V = \; -n + n^2 (1-x)^2 \sum_{i=0}^\infty x^{2i}
, \quad {\mathrm{where}} \;\; x = \left(\frac{1}{2}\right)^\frac{1}{k}. 
\end{equation}
\noindent When either $k$ or $n$ is small, this formula does not agree with our simulation results,
but the match is so close when $k$ is large and $n >> k$ that we continue
the derivation by summing the series in (\ref{conjectured-hip-variance}).
{\large
\begin{eqnarray}
V/n^2 
= & \frac{(1-x)^2}{(1-x^2)} - \frac{1}{n} \\
         < & \frac{(1-x)^2}{(1-x^2)}
         \;\;=\;\;   \frac{1}{2} \left(\frac{1}{x} - 1 \right) - \frac{(x-1)^2}{2x(x+1)} \\
         < & \frac{1}{2} \left(\frac{1}{x} - 1 \right) 
         \;\;=\;\;  \frac{1}{2} \left(2^\frac{1}{k} - 1 \right) 
         \;\;=\;\;  \frac{1}{2} \left(e^\frac{\ln 2}{k}  - 1 \right), \\
\approx & \frac{1}{2} \; \frac{\ln 2}{k}.
\end{eqnarray}}
\noindent Then, because HIP estimators are unbiased,
\begin{equation} \label{fm85-hip-rmse}
\mathrm{RMSE}(\hat{N}_{\mathrm{HIP}})/n 
\;\;\approx\;\; \sqrt{\frac{\ln 2}{2k}}
\;\;\approx\;\; \frac{0.58870501}{\sqrt{k}}.
\end{equation}
\noindent Because the constant in (\ref{fm85-hip-rmse}) matches our simulations to 6 decimal digits 
(the noise floor of the empirical measurements), 
we conjecture that a more rigorous analysis of the FM85 HIP estimator would arrive at essentially the same result.

\section{MDL Estimator for FM85 Sketches}

We have not analyzed the error of the Minimum Description Length estimator for either FM85 or HLL, but our
simulations do provide approximate values for the leading constants of their error 
formulas.\footnote{The paper \cite{maxlikepaper}
proposed and evaluated a Maximum Likelihood estimator for HLL. Judging from the
paper's plots, the results were very similar to our MDL results for HLL. This isn't surprising
given the close connection between the MDL and Maximum Likelihood paradigms.}
It is interesting to compare these constants with those of the best summary statistic estimators for the two
types of sketch:

\begin{center}{\footnotesize
\begin{tabular}{|c|c|c|}
\hline
  & HLL Sketch & FM85 Sketch \\
\hline
Summary Statistic Estimator    & 1.039 & 0.693 \\
MDL Estimator & 1.037 & 0.649 \\
\hline
\end{tabular}}
\end{center}

Apparently, FM85 sketches benefit more from the MDL paradigm than HLL sketches do.
We speculate that the lossy mapping from an FM85 sketch to an HLL sketch
discards most of the extra information (beyond the summary statistic) that an MDL 
estimator would be able to exploit.

\section{Simulations}

\begin{figure}
\begin{center}
\includegraphics[width=0.48\linewidth]{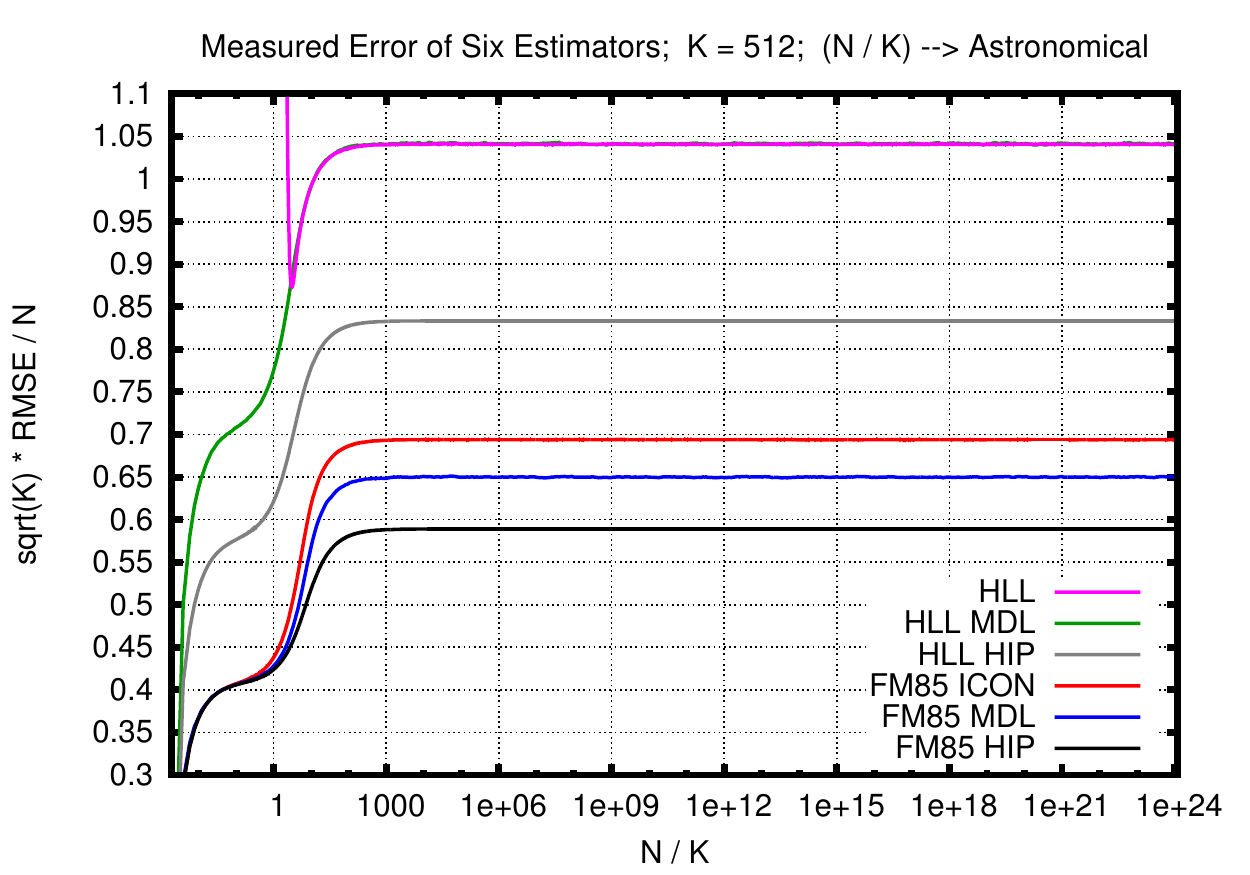} \quad 
\includegraphics[width=0.48\textwidth]{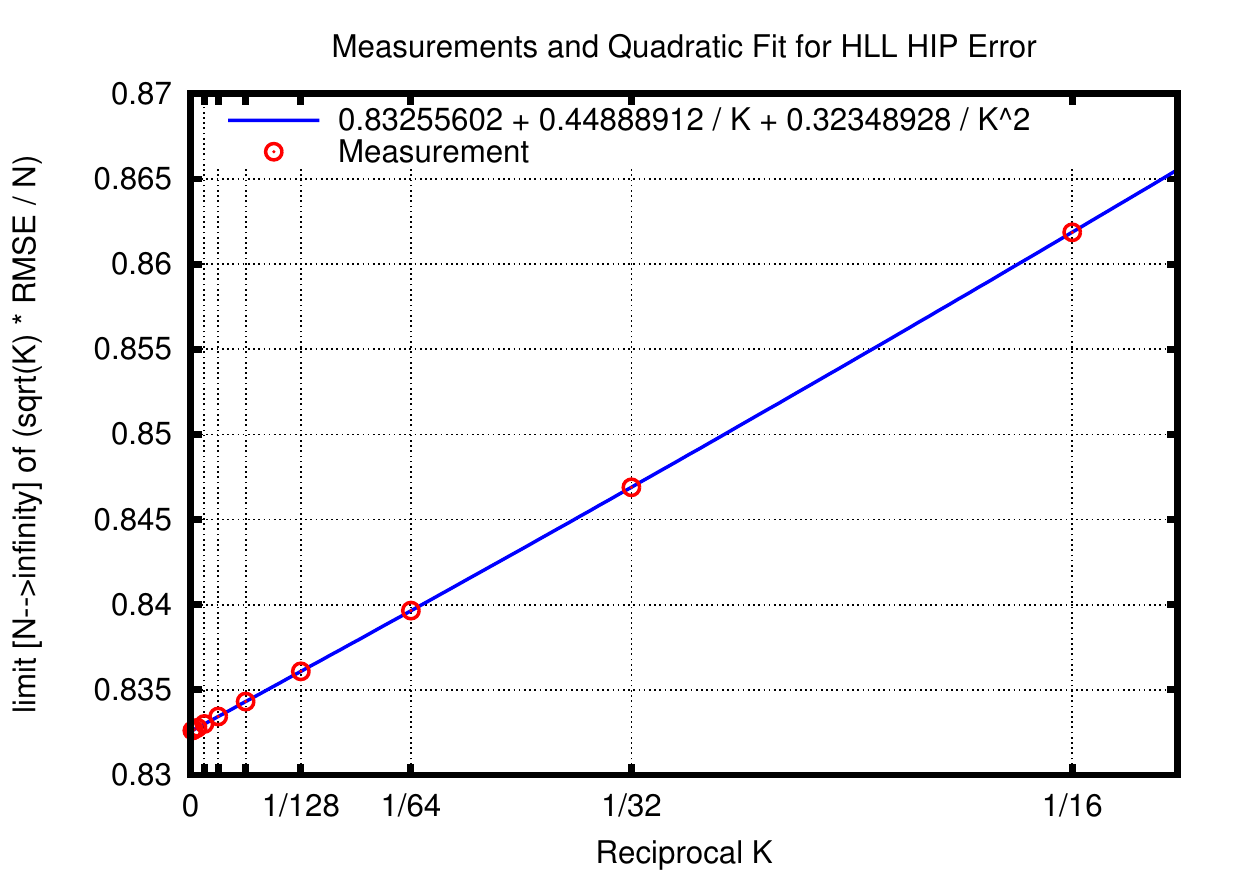}
\caption{These plots illustrate a two-stage procedure for empirically measuring a 
double asymptotic limit.}
\label{nutshell-figure}
\end{center}
\end{figure}

In this section we use Monte Carlo methods to approximately measure
the Error Constant for each of the six estimators that are 
the subject of this paper. Recall that
\begin{equation}
\mathrm{Error\;Constant} = \lim_{k\rightarrow\infty} (\lim_{n\rightarrow\infty} (\sqrt{k} \times \mathrm{RMSE} / n)).
\end{equation}
Evaluating a double limit empirically is not a simple task, but in this case
it can be done with the following two-stage method.
First, we employ the exponentially accelerated simulator for coupon collection
that is described in Appendix B.
This can simulate streams whose length is literally astronomical
($10^{24}$ is roughly the number of stars in the universe) at a cost of
only $O(x \log x)$, where $x$ is the number of collectible
coupons. As can be seen in Figure~\ref{nutshell-figure}(left),
the quantity 
$\sqrt{k} \times \mathrm{RMSE} / n$ effectively reaches its
infinite-$n$ limit long before that, and except for the usual tiny oscillations,
a measurement anywhere along the flat part of the curve
is a noisy estimate of the desired number. To reduce the noise,
we make hundreds of measurements along the flat part 
(between $n=k\cdot2^{25}$ and $n=k\cdot2^{75}$) for each stream, and average them. 

We can now estimate $\lim_{n\rightarrow\infty} (\sqrt{k} \times \mathrm{RMSE} / n)$
for any fixed value of $k$, but we still need to take the limit as $k$ goes to infinity.
Our technique for doing this is model-based extrapolation. 
Theorem 1 in \cite{hllpaper} provides a formula for the Standard Error of 
the HLL estimator that has the form
$\lim_{n\rightarrow\infty}(\mathrm{RMSE} / n) = (c_0/\sqrt{k}) \cdot \mathcal{F}(1/k) + \mathrm{(tiny\;oscillation)} + o(1)$.
$\mathcal{F}()$ is a rational function of $(1/k)$ that converges to 1 as $(1/k)$ goes to zero.
Our model ignores the last two terms (which are both tiny), and replaces $\mathcal{F}()$
with a quadratic of the form $(1 + a/k + b/k^2)$, so after multiplying through we have

\begin{equation}
\lim_{n\rightarrow\infty} (\sqrt{k} \times \mathrm{RMSE} / n)
\;\approx\;  c_0 + c_1/k + c_2/k^2.
\end{equation}

\noindent The three constants can be estimated by first
measuring $\lim_{n\rightarrow\infty} (\sqrt{k} \times \mathrm{RMSE} / n)$ at
several small-to-moderate values of $k$, then calculating a least-squares
fit of a quadratic in $(1/k)$ to the measurements. The value of $c_0$
is the desired estimate of the Error Constant. 
We validated this methodology using the theoretical values
of $c_0$ and $c_1$ for the HLL HIP estimator that were derived by \cite{tingHIP}
We used our accelerated simulator to estimate $\lim_{n\rightarrow\infty} (\sqrt{k} \times \mathrm{RMSE} / n)$
for $k$ in $\{2^4, 2^5, \ldots 2^{13}\}$, then obtained the quadratic fit
shown in Figure~\ref{nutshell-figure}(right), which illustrates
how the extrapolation to $k=\infty$ has been converted
into a short-range extrapolation to $(1/k)=0$.
Here are the results of the validation experiment:
\begin{center}{\footnotesize
\begin{tabular}{|c|c|c|}
\hline
  & $c_0$ & $c_1$ \\
\hline
Theoretical & 0.83255461 & 0.449347 \\
Estimated   & 0.83255602 & 0.448889 \\
\hline
\end{tabular}}
\end{center}
To guard against reporting spurious (i.e.~coincidental) 
levels of agreement between our estimates and the corresponding
theoretical values, the following table contains a column
labeled D1 which is a rough measure of the uncertainty of each empirical
estimate; it was generated by repeated subsampling from the
hundreds of measurements along the ``flat part'' of each error curve,
and is stated as a number of digits beyond which the estimate is probably
noise. The column labeled D2 indicates the number of digits that match between the 
measurement and the reference value 
\begin{center}{\footnotesize
\begin{tabular}{|c|c|c|c|c|c|c|}
\hline
Error Constant & Estimate & D1 & D2 & \multicolumn{3}{|c|}{Reference Value} \\
\hline
HLL       & 1.0390092 & 4.4 & 4.4 & 1.03896176 & $\sqrt{3(\ln 2)-1}$ & \cite{hllpaper} \\ 
HLL HIP   & 0.8325560 & 5.9 & 5.8 & 0.83255461 & $\sqrt{\ln 2}$      & \cite{tingHIP} \\ 
\hline
FM85 ICON & 0.6931697 & 4.5 & 4.5 & 0.69314718 & $ \ln 2$ & \\            
FM85 HIP  & 0.5887044 & 5.9 & 6.0 & 0.58870501 & $\sqrt{(\ln 2)/2}$ & \\  
\hline
HLL MDL  & 1.036624 & 3.5 & & & & \\
FM85 MDL & 0.649057 & 3.5 & & & & \\
\hline
\end{tabular}}

\end{center}
\noindent We can also define $\mathrm{Bias\;Constant} = \lim_{k\rightarrow\infty} (\lim_{n\rightarrow\infty} (k \times \mathrm{Bias} / n))$
and measure it using a similar procedure.
[The measured biases of the HLL, HLL HIP, and FM85 HIP estimators are all nearly zero, as predicted by theory.]

\begin{center}{\footnotesize
\begin{tabular}{|c|c|c|c|c|c|}
\hline
Bias Constant & Estimate & D1 & D2 & \multicolumn{2}{|c|}{Reference Value} \\
\hline
FM85 ICON  & 0.24028 & 3.2 & 3.6 & 0.24022651 & $(\ln 2)^2/2$ \\
FM85 MDL   & 0.30685 & 1.9 & & & \\
HLL MDL    & 1.00760 & 2.3 & & & \\
\hline
\end{tabular}}
\end{center}

\noindent This section's results are all for the infinite-$n$ limit, but
Appendix C provides some intuition for why the small-$n$ 
Standard Error of the FM85 estimators is roughly $0.408/{\sqrt{k}}$,
as can be seen in Figure~\ref{nutshell-figure}(left).

\section{Compression Techniques for FM85 Sketches}

It has been proved that arithmetic coding \cite{wittencoding}
can (on average, 
with an overhead of 2 bits)
compress the state of any stochastic process down to its entropy,
which in the case of FM85 sketches is 4.70 bits per row. 
Because arithmetic coding is too slow for high-performance systems, we mention
that the columns of an FM85 sketch are essentially Bloom
Filters that can be quickly compressed to
nearly their entropy by employing the technique that is used for
postings lists in document retrieval systems, namely encoding the
gaps between successive 1's using a Golomb code.  
A better idea is to compress the sketch's 8
highest-entropy columns (viewed as $k$ bytes) using pre-constructed Huffman
codes. The remaining columns, which contain a total of roughly $k/30$
``surprising'' matrix bits, can be handled using the Bloom filter
technique. With careful programming, this can all be accomplished during a single pass
at a cost of $O(k)$ independent of the number of columns. 
Preliminary calculations show that this scheme would compress 
the sketches to about 4.9 bits per  row. 
However, the zstd compression library \cite{zstdlibrary}
can compress
FM85 sketches to 5.2 bits per row. Because 5.2 / 2 = 2.6 $\;<\;$ 2.83, this is already
good enough to beat the space efficiency of any possible implementation of
compressed HLL.


\subsection{Details}
The above figure of 5.2 bits per row can be achieved as follows. First, the 
($k$ $\times$ $\infty$) matrix of indicator bits is represented by an offset and
a ($k$ $\times$ 32) sliding window.
All coupons to the left of the sliding window
have already been collected. Almost no coupons to the right of the sliding window
have been collected, but if any have been, they are handled separately. 
Next, the 32 in-window bits from each row are interpreted as a 32-bit integer,
then subjected to a conditional rotation (see below), then re-interpreted as 4 bytes.

The resulting ($k$ $\times$ 4) matrix of bytes is transposed from row-major to column-major
order in memory, then fed into the zstd compression library, specifying compression-level $=$ 1.
The column-major order is important because it brings matching byte patterns closer together, 
which helps an LZ77 compressor like zstd to run faster and achieve better compression.  
Now we explain the conditional rotation: when the number of collected coupons
exceeds $3.3 \times k$, the 32 bits from each row of the sliding window are rotated
left by one position before being split into bytes.\footnote{In our actual code,
the low order bit of a 32-bit integer represents the leftmost column of the window,
so the physical rotation is to the right.} This affects the compression because
the columns have different entropies, and rotating them causes 
different groups of 8 columns to be packaged together for input into zstd.

\begin{figure}[t]
\begin{center}
\includegraphics[width=0.45\linewidth]{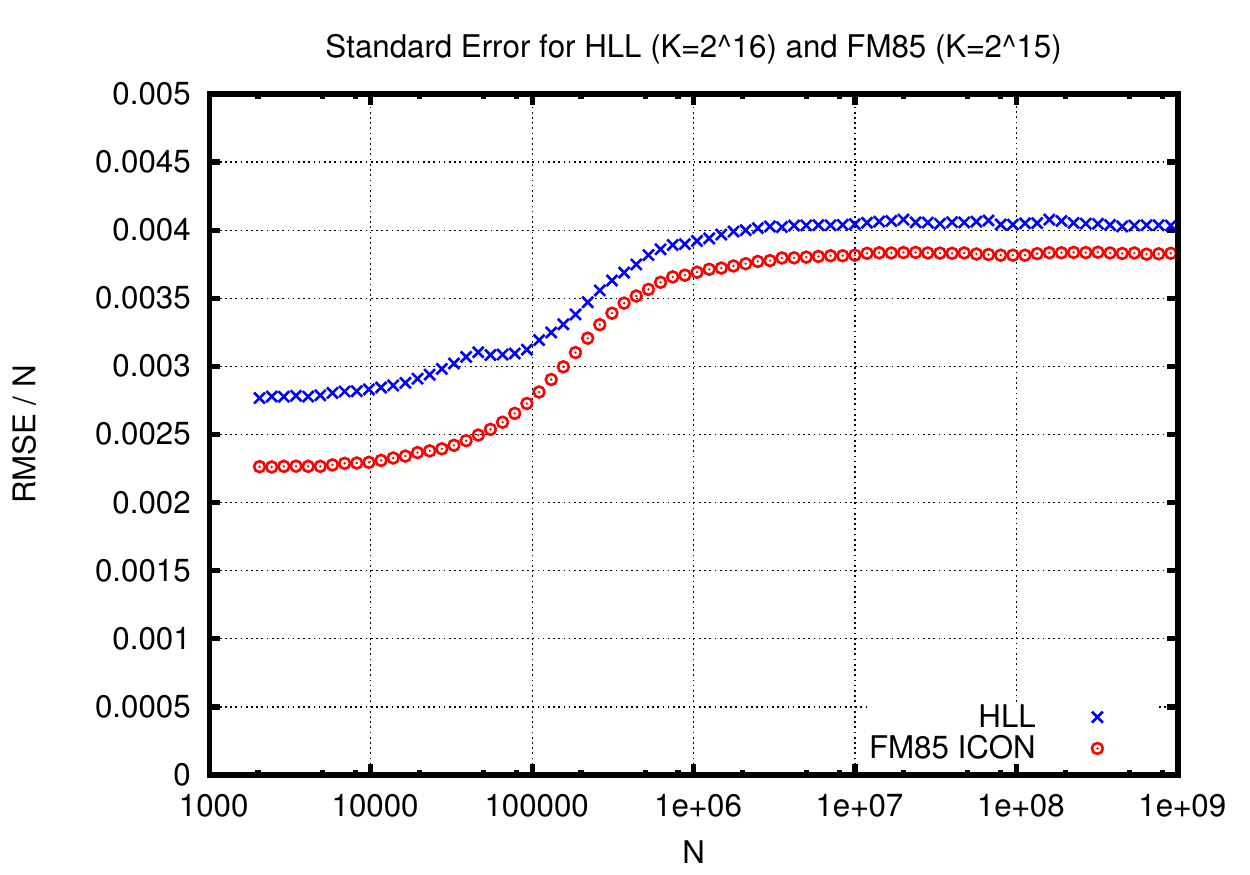} \quad
\includegraphics[width=0.45\textwidth]{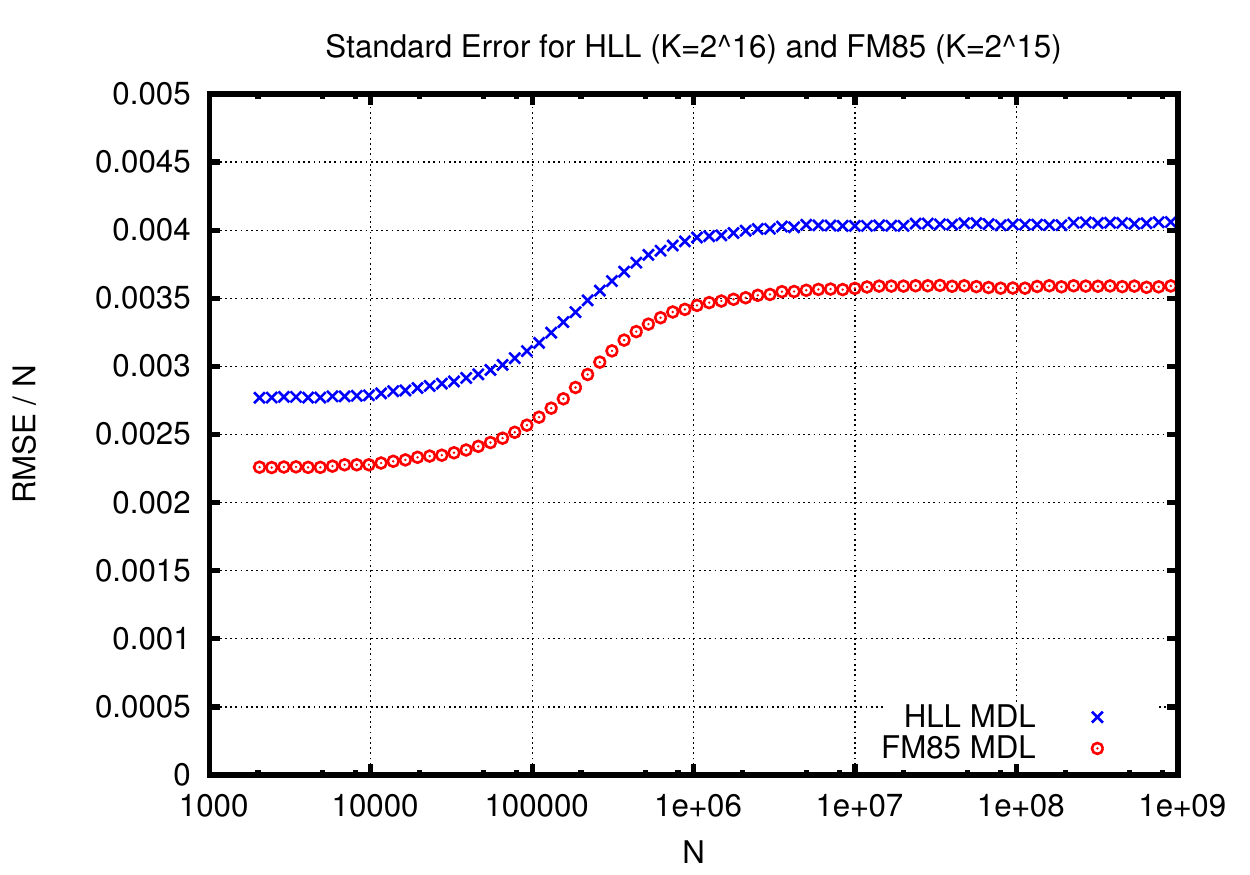}
\includegraphics[width=0.45\textwidth]{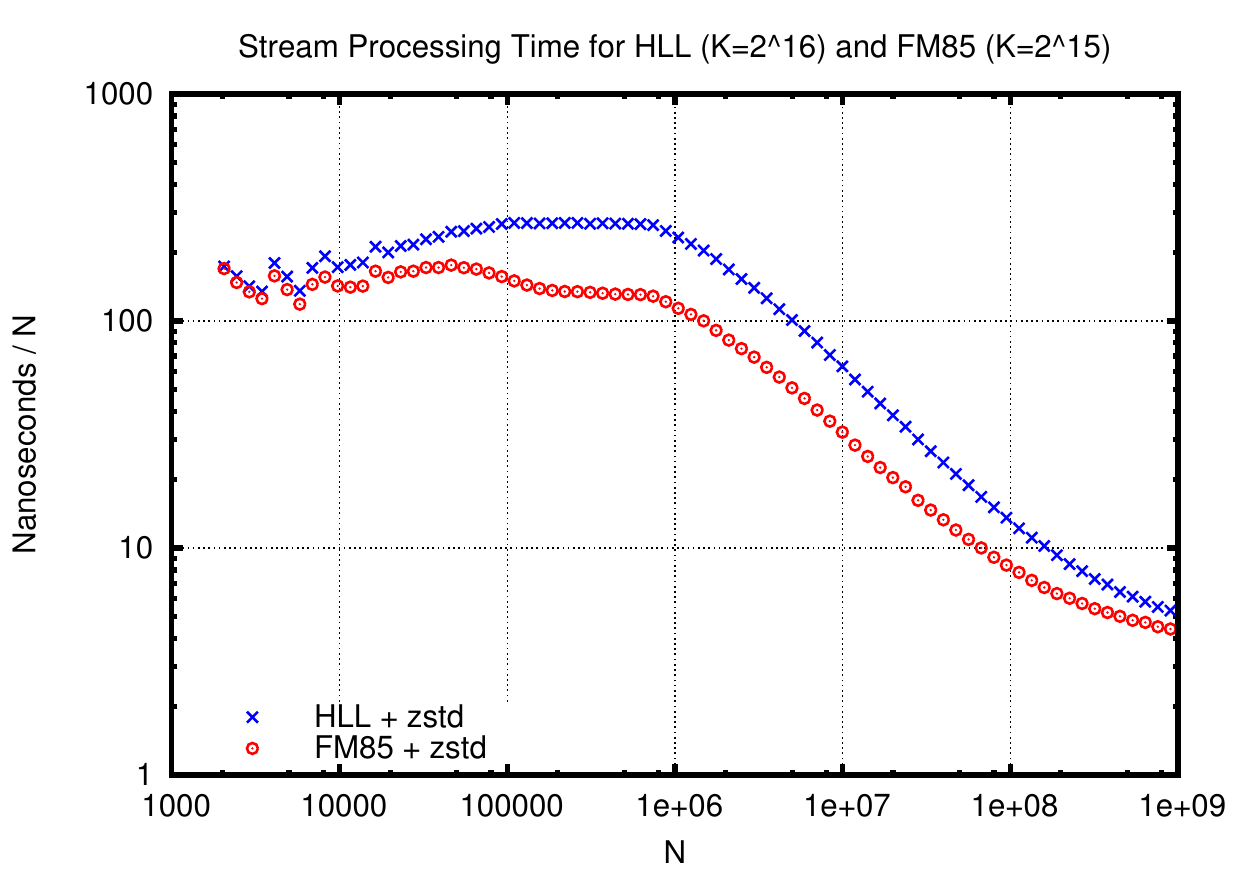} \quad 
\includegraphics[width=0.45\linewidth]{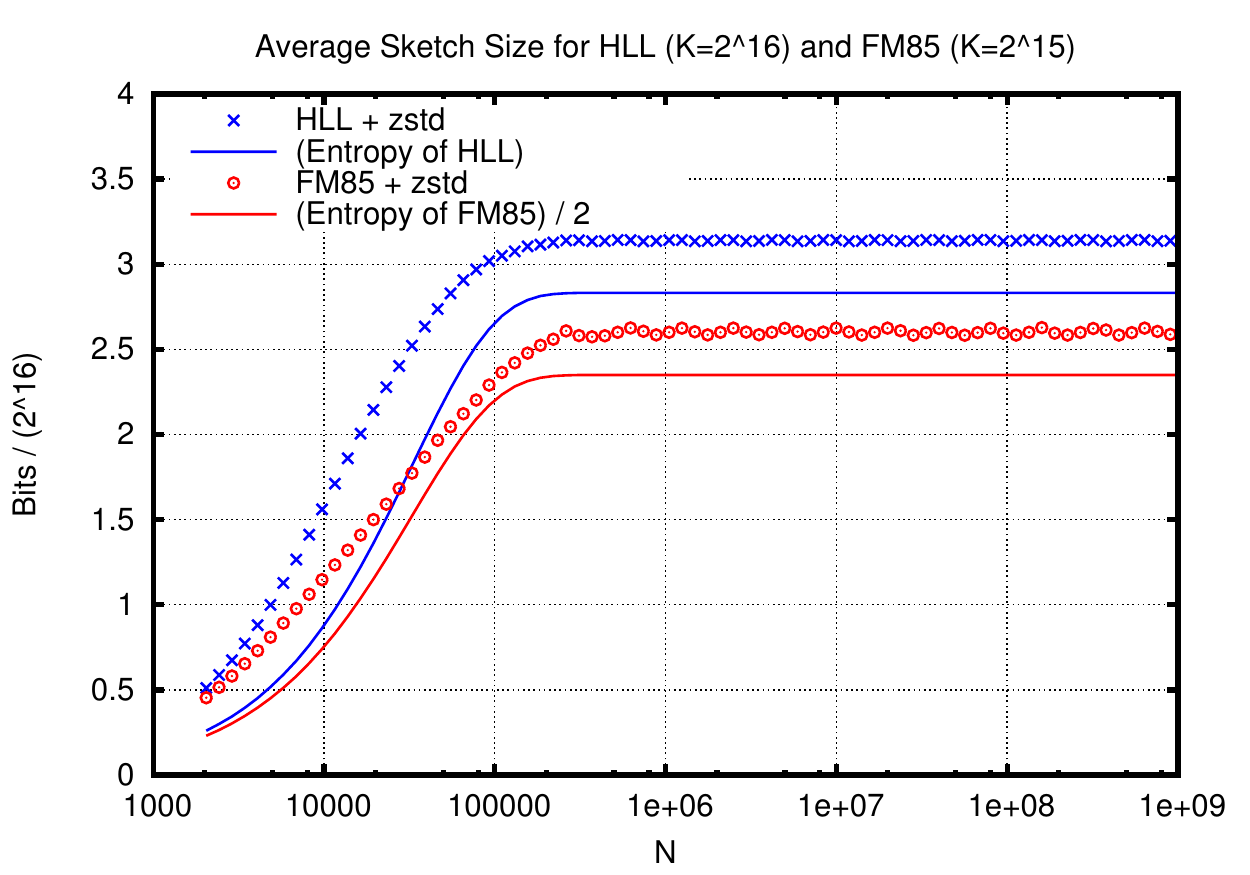}
\caption{Compressed FM85 sketches are smaller than the {\em entropy} of HLL sketches that have worse accuracy,
and are faster than an HLL implementation that is written in the same style using the same compression technology.}
\label{winning-figure}
\end{center}
\end{figure}

\section{Experimental Evaluation}\label{prototype-section}

Our prototype implementation of compressed FM85 uses the zstd library \cite{zstdlibrary}
and the sliding window technique described in the previous section. Because a 
comparison against an existing HLL implementation would be affected by numerous
design choices that were made differently between the two programs, we 
wrote a new implementation of compressed HLL based on the zstd library.
As much as possible, we did things the same way in both programs and 
used equivalent parameters.

Both programs are written in C and use an update buffer that can hold 2000 items. When the
buffer is full, the sketch is uncompressed, the updates are processed, then 
the sketch is recompressed. Both programs track the leftmost interesting column
of the coupon tableau. As $n$ gets larger, an increasing fraction of updates
can be discarded instead of entering the buffer, which greatly speeds up
the algorithm because the sketch doesn't need to be uncompressed as often.
In both programs, CityHash is used to map each item to a pair of 64-bit hashes. Leading zeros are counted
in one of them to obtain the column index, while the row index is determined by the
low bits of the other hash. 

The HLL program's register array is simply an array of $k$ bytes that is fed 
straight into the zstd compression library (at compression-level $=$ 1);
unlike with FM85, no fancy tricks are required.
The FM85 estimators are as described in this paper. The HLL HIP estimator is 
as described in \cite{tingHIP}. The HLL estimator is similar to the one in \cite{heule2013hll}.
The HLL MDL estimator is related to the Maximum Likelihood estimator in \cite{maxlikepaper}.

Because the Standard Errors of the HLL and FM85 HIP estimators are respectively
$\sqrt{(\ln 2)/k}$ and $\sqrt{(\ln 2)/2k}$, their HIP accuracy will be exactly the
same if HLL is allowed to use a value of $k$ that is twice as big. That is
why we specify $k=2^{15}$ for FM85 and $k=2^{16}$ for HLL.

The results of this experiment are shown in Figure~\ref{winning-figure}. Each estimator
is compared against the other sketch's estimator from the same category. Clearly, the
FM85 ICON estimator is more accurate than the HLL estimator, and the FM85 
MDL estimator is more accurate than the HLL MDL estimator.
Not shown are the accuracies of the two HIP estimators, which are equal, as per our experimental 
design.

The time plot shows that both algorithms speed up with increasing $n$, but FM85 is always faster.
This surprised us, because FM85 is sending twice as many bytes to zstd; apparently the fact
that they are more compressible allows zstd to run faster.

The space plot
shows that zstd compresses both types of sketch to slightly above 
their entropy, but FM85 is always 
smaller.\footnote{This plot shows the final sizes of the sketches;
while the stream is being processed there is also a buffer for updates that raises both
curves by $0.6714 = 2000 \times (16 + 6) / 65536$.}
In fact, the compressed FM85 sketches are smaller
than the entropy of HLL, so no possible implementation of HLL using the
currently known estimators can match FM85's space efficiency.
Furthermore, because the powerful MDL technique barely improves on the accuracy
of the original HLL estimator,
we conjecture that no HLL implementation using {\em any} estimator
can match FM85's space efficiency.

\section{Related Work}

Aside from HLL \cite{hllpaper}, the other cardinality estimation sketch that is in widespread 
today is KMV \cite{bar2002counting,beyer2009distinct,cohen2007summarizing,thetaICDT,giroire2009order},
which achieves a Standard Error of $1 / \sqrt{k-2}$ by tracking the
$k$ numerically smallest 64-bit hashes that have been seen so far.
Although KMV sketches consume 10 times as much memory 
as HLL sketches, they are more accurate for set intersections, 
and it has been argued that KMV provides more operational 
flexibility in large-scale systems environments.

\noindent The HLL family of sketches began with the FM85 paper \cite{fm85paper},
which proposed 
an estimator based on the average position of the leftmost uncollected coupon in each row.
Because the sketch required $k$ words of memory, the authors
speculated that a lossy 8-column sliding window into the coupon
tableau might suffice.
With the benefit of hindsight we know that this specific
idea didn't pan out,
but this shows that at the very beginning of the field it was
understood that sketches can be smaller than naive implementations
would suggest.

\vspace{0.5em} \noindent In the ``LogLog'' papers that followed, the authors switched to
estimators based on the average position of the
rightmost collected coupon in each row, allowing
the sketch to shrink to $6 k$ bits. The estimator in the
HyperLogLog paper \cite{hllpaper}
employed a harmonic average, and spliced in the bitmap estimator from 
\cite{whangpaper} for the small-$n$ regime. 
There was still a spike in error at the crossover point, which the 
HLL++ \cite{heule2013hll} implementation reduced to a small bump 
by applying an empirical bias correction. 
\cite{maxlikepaper} showed that a Maximum Likelihood estimator
yields a similar error curve which lacks the bump
[compare the blue curves in top two plots of the current paper's 
Figure~\ref{winning-figure}.]

\vspace{0.5em} \noindent HIP estimators were discovered independently by \cite{cohenHIP}
and \cite{tingHIP}, and both papers used HLL sketches to illustrate how this idea can be applied.

\vspace{0.5em} \noindent \cite{frenchthesis} proved an upper bound of 3.01 bits per row for the entropy of HLL.

\vspace{0.5em} \noindent Several practical implementations of HLL have employed data compression, 
including \cite{heule2013hll} and \cite{datasketcheslink}.

\vspace{0.5em} \noindent Finally, the KNW sketch \cite{knwpaper} is space-optimal in a theoretical sense,
but the constant factors are unknown, and to the best of our knowledge
it has never been used in a real system.

\section{Conclusion}

We have shown that compressed FM85 with our new estimators has a 
space / accuracy tradeoff that cannot be matched by any 
implementation of compressed HLL that uses the currently
known estimators. Compressed FM85 can also be fast; our prototype
can process a stream of 1 billion items in 4.3 seconds. We anticipate
that this algorithm will see production use at companies and government
agencies that require space efficiency in their cardinality estimation systems.

\section*{Appendix A: Variance of the Random Variable C}

\paragraph{Theorem:}
Let $C$ be the number of collected coupons in an FM85 sketch,
and $M$ a large but finite number of columns. Then for sufficiently large $k$
with $n >> k$ and $k \cdot 2^M >> n$, 
$\;\;\sigma^2(C) \; \approx \; k$.
\begin{proof}
\noindent Recall that $\sigma^2(C) = k \cdot S$, where $S = \sum_{j=1}^M q^n_j (1 - q^n_j)$,
and $q^n_j = 1 - (1 - \frac{1}{k 2^j})^n$. Then:
\begin{eqnarray}
S = & 
\sum_{j=1}^M \left[ 
\left(1 - (1 - \frac{1}{k 2^j})^n\right) \;-\;
\left(1 - (1 - \frac{1}{k 2^j})^{2n}\right) 
\right] \label{app-1-b} \\
\approx &
\sum_{j=1}^M \left[ 
\left(\frac{1}{e}\right)^{\frac{n}{k2^j}} -\;
\left(\frac{1}{e}\right)^{\frac{n}{k2^{j-1}}}
\right]
\label{app-1-c} \\
= & 
\left(\frac{1}{e}\right)^{\frac{n}{k2^M}} \;-\;\;
\left(\frac{1}{e}\right)^{\frac{n}{k}}
\label{app-1-d} \\
\approx & 1 \;-\; 0. \label{app-1-e}
\end{eqnarray}
\noindent The approximations in (\ref{app-1-c}) are valid for sufficiently large $k$.
The sum in (\ref{app-1-c}) telescopes, resulting in (\ref{app-1-d}).
The two approximations in (\ref{app-1-e}) are valid
because $k \cdot 2^M >> n$, and $n >> k$.
Therefore $\sigma^2(C) \;=\;k \cdot S \;\approx\; k \cdot 1$.
\end{proof}

\section*{Appendix B: Exponentially Accelerated Simulation of FM85} %

We fix the number of columns
at 96, then instantiate the complete set of $96 \times k$ coupons.
The coupon probabilities $1/(k \cdot 2^{j})$ induce a probability distribution over
coupon discovery sequences. We draw specific sequences from that
distribution using the ``exponential clocks'' method \cite{Bollobas}.
Each discovery sequence defines a sequence of waiting-time distributions
for the successive coupons. These distributions are
geometric, and are parameterized by the total amount
of uncollected probability that remains right
before each novel coupon is encountered.
By summing a sequence of draws from these waiting-time distributions, we
obtain the value of $n$ at which each coupon is collected. This technique
allows us to simulate streams of length roughly $k \cdot 2^{96}$, but 
to avoid edge effects at the right side of the coupon matrix, we stop
at $n = k \cdot 2^{80}$. Because HLL estimators can be applied to
FM85 sketches, we evaluate all six estimators on each simulated stream. 
This not only saves on CPU time, it
gives the measurements more power to discriminate between the estimators.

\section*{Appendix C: The Accuracy of FM85 when N $<$ K$^{2/3}$}

TSBM (an acronym for ``triple-size bitmap'') was the first novel
estimator that we devised for FM85 sketches:
\begin{equation}
\hat{N}_{\mathrm{TSBM}} = 3k \; (H(3k) - H(3k-C)),
\end{equation}
\noindent where $H(i)$ denotes the  $i$'th Harmonic Number. 
Although this estimator only applies to the small-$n$ regime and 
has been superceded by the more rigorous ICON estimator,
the idea that it was based on is worth discussing because
it provides an intuitive explanation for the fact that the small-$n$
accuracy of FM85 is roughly a factor of $\sqrt{3}$ better than that of
HLL.

It will be convenient to refer to collected coupons as balls, 
and collectible coupons as bins. When two balls land in the same
row of an FM85 sketch, the probability of them landing in the
same bin is
\begin{equation}\label{triple-fact}
\sum_{j=1}^{\infty} 2^{-2i} = 1/3,
\end{equation}
\noindent while the probability of them landing in different bins is $1 - 1/3 = 2/3$.

Now consider a triple-size bit map which has $k$ rows and 3 coupons
in each row, all equiprobable. When two balls land in the same
row of this kind of sketch, the probabilities of them landing in the
same bin or different bins are once again $1/3$ and $2/3$.

Keeping this mathematical coincidence in mind, consider a pair of 
synchronized runs of FM85 and TSBM. Until the first row-level collision occurs, the probability
of a coupon-level collision occurring on the next draw will be identical for the two sketches.
According to the birthday paradox, this typically won't happen until roughly 
$n=\sqrt{k}$, so averaging over all possible runs,
the two sketch's mappings from $n$ to $E(C)$ should be nearly the same when 
$n < \sqrt{k}$, which implies that their ICON estimators should be nearly the same. 


We now point out that 
ICON estimators are closely related to estimators
that map $C$ to the expected discovery time of the $C$'th
coupon. When the coupons are equiprobable (as in a bitmap sketch), the
expected discovery time can be written in a closed form
\begin{equation}
\hat{N}_{\mathrm{bitmap}} = k \; (H(k) - H(k-C)).\quad \quad \cite{tingHIP}
\end{equation}

\vspace{0.5em}
\noindent \cite{whangpaper} showed that this estimator and its variance can
be approximated by
\begin{eqnarray}
\hat{N}_{\mathrm{bitmap}} \approx & k \; \ln(k/(k-C)). \\
\sigma^2(\hat{N}_{\mathrm{bitmap}}) \approx & k \; e^{n/k} - n - k. \label{whang-variance}
\end{eqnarray}

\noindent When $n << k$, (\ref{whang-variance}) can be further approximated by replacing
$e^{n/k}$ with the first three terms of its Maclaurin series:

\begin{equation}
\sigma^2(\hat{N}_{\mathrm{bitmap}}) \;\approx \;
k [1 + n/k + n^2/(2k^2)] - n - k \;= \;
n^2/(2k).
\end{equation}

Clearly, the variance decreases by a factor of 3 when k is increased by a factor of 3,
which means that the Standard Error of a triple-size bitmap is a factor of $\sqrt{3}$ 
smaller than that of an ordinary bitmap.
Recall that HLL uses the bitmap estimator when $n < k$, while we have just argued that 
the small-$n$ behavior of the FM85 coupon collection process is 
similar to that of a triple-size bitmap. 
Therefore the small-$n$ Standard Error for FM85 should be about
a factor of $\sqrt{3}$ lower than that of HLL.\footnote{In more detail, the
small-$n$ Standard Errors should be roughly
$1/\sqrt{2k} = 0.707/{\sqrt{k}}$ for HLL, and 
$1/\sqrt{6k} = 0.408/{\sqrt{k}}$ for FM85.
Empirical measurements yield similar values, as can be seen later in this section.}

\paragraph{Extending the argument to N $<$ K$^{2/3}$:} 
When 3 balls land in the same row of an FM85 sketch, the probabilities
of them ending up in 1, 2, or 3 different bins are respectively
$1/7, 4/7$, and $2/7$. The corresponding probabilities for a triple-size bitmap are
$1/9, 6/9$, and $2/9$. These are not equal to the FM85 probabilities,  but they
are fairly close.
Now consider a pair of synchronized runs of FM85 and TSBM. As long as every row contains
at most 2 collected coupons, the probability of a coupon-level collision occuring on the next draw 
is similar for the two sketches.\footnote {Especially because most rows still contain 
either 0 or 1 collected coupons, and it is only the uncommon 2-coupon rows 
that are adding their slightly different collision probabilities into the total.}
According to a generalization of the Birthday Paradox,
this condition will usually be satisfied while $n < k^{2/3}$.

\paragraph{Experimental Results:}

The error curves in Figure~\ref{nutshell-figure} (left) were generated with $k = 512$, so $k^{2/3}=64$.
At $n = 64$, the ratio (HLL Standard Error) / (FM85 ICON Standard Error) 
is\footnote{These aren't raw Standard Errors; both the numerator and denominator 
of the ratio have been scaled up by $\sqrt{512}$.} 
$0.716768 / 0.408845 = 1.753 \approx \sqrt{3}$.
For the algorithms' MDL estimators, the ratio is
$0.710784 / 0.407660 = 1.744 \approx \sqrt{3}$.
However, the ratio for their HIP estimators is
$0.578730 / 0.407170 = 1.4213 \approx \sqrt{2}$.

In the experiment that generated figure~\ref{winning-figure}, HLL was run with a value of
$k$ that was twice the value used with FM85; that is why the small-$n$ ratio 
(HLL Standard Error) / (FM85 Standard Error) is $\sqrt{3/2}$ for the
non-HIP estimators, and unity for the HIP estimators.

\section*{Appendix D: The Data Sketches Library}

Data Sketches \cite{datasketcheslink} is an open-source library of sketching implementations
that as of August 2017 is being used by several internet and big-data companies.
Despite being written in Java, the library is fast because of a Fortran-like programming style
that focuses on arrays of primitive types, and also because of strategies for avoiding 
garbage collection that were devised by Lee Rhodes, the architect of the Data Sketches 
library, and Eric Tschetter, the architect of the Druid column store \cite{druidlink}.

Currently, the library doesn't provide a full-fledged implementation of compressed FM85, but its implementation of HLL
includes several details that were motivated by the research reported in this paper.

For example, when $n < k / 10$, the sketch is actually FM85 rather than HLL, and employs either 
the FM85 ICON estimator or the FM85 HIP estimator. As can be seen in Figure~\ref{nutshell-figure} (left),
these two estimators have roughly the same Standard Error when $n < k / 10$. As discussed
in Appendix C, this FM85 error is a factor of $\sqrt{3}$ lower
than that of the HLL estimator, and
a factor of $\sqrt{2}$ lower than that of the HLL HIP estimator.

When $n$ reaches $k/10$, the library converts the FM85 sketch into an HLL
sketch by discarding
all information about cells that are to the left of the rightmost collected
coupon in each row.
The HIP estimation scheme handles this mid-stream change of
sketching algorithm by overwriting $R$ with the HLL amount of remaining probability,
which is different from the FM85 amount of remaining probability.
The accumulator $A$ isn't touched, and although its error will eventually grow to that
of the HLL HIP estimator, this is a gradual process rather than a sudden one,
so the superior accuracy of FM85 persists for a while.

After the transition, the library stores the HLL sketch in just over
4 bits per row by using an offset and an array of nybbles. 15 of the possible
values are interpreted by adding the offset, while the $16^{\mathrm{th}}$ value 
tells the algorithm to look in a hashmap of exceptions.
The average number of exceptions varies with $k$ but is always much smaller.
For example, when $K=2^{12}$, the average number of exceptions is 2.2.

It should be mentioned that the apples-to-apples comparison between
Compressed FM85 and Compressed HLL in Section~\ref{prototype-section} of this
paper employed a specially-written implementation of HLL, not the Data Sketches
implementation.

\end{document}